\documentclass{aa}  

\usepackage{graphicx}
\usepackage{txfonts}
\usepackage{stmaryrd}
\usepackage{caption}
\usepackage{subcaption}

\newcommand{\deriv}[2]{\frac{d{#1}}{d{#2}}}
\newcommand{\pderiv}[2]{\frac{\partial{#1}}{\partial{#2}}}
\newcommand{\diff}[1]{\left\llbracket{#1}\right\rrbracket}

\begin{document}

   \title{Magnetic Rayleigh-Taylor Instability at a Contact Discontinuity With Oblique Magnetic Field}
   \titlerunning{Magnetic Rayleigh-Taylor Instability at a Contact Discontinuity}


   \author{E. Vickers \inst{1},
   			I. Ballai \inst{1}
          \and
          R. Erd\'{e}lyi \inst{1,}\inst{2}
          }

   \institute{School of Mathematics and Statistics, University of Sheffield, Hicks Building, Hounsfield Rd., Sheffield, UK
         \and
	 Dept. of Astronomy, E\"otv\"os L. University, P\'azm\'any P. s\'et\'any 1/A, Budapest, H1117, Hungary \\
              \email{evickersl@sheffield.ac.uk} }

   \date{Received 9 August 2019; Accepted 24 November 2019}

 
  \abstract
   {}
   {In the present work we investigate the nature of the magnetic Rayleigh-Taylor instability at a density interface permeated by an oblique, homogeneous magnetic field in an incompressible limit.}
   {Using the system of linearised ideal incompressible magnetohydrodynamics (MHD) equations we derive the dispersion relation for perturbations of the contact discontinuity by imposing the necessary continuity conditions at the interface. The imaginary part of the frequency describes the growth rate of waves due to instability. The growth rate of waves is studied by solving numerically the dispersion relation.}
   {The critical wavenumber at which waves become unstable, present for a parallel magnetic field, disappears, due to the inclination of the magnetic field and instead waves are shown to be unstable for all wavenumbers. Theoretical results are applied to diagnose the structure of the magnetic field in prominence threads. When applying our theoretical results to observed waves in prominence plumes, we obtain a wide range of field inclination angle; from 0.5$^\circ$ up to 30$^\circ$. These results highlight the diagnostic possibility of our study. 
}
   {}


   \maketitle
%

\section{Introduction}

The hydrodynamic Rayleigh-Taylor instability (RTI), first explored by \cite{rayleigh} and \cite{taylor}, for the incompressible case, concerns the instability of an interface, separating two fluids with different properties, where a dense fluid is supported above a lighter fluid, in the presence of a transversal gravitational force (or equivalently an acceleration of the fluid system in the direction of the denser fluid). The properties and the consequences of this instability are well established. The possibility of a similar instability in magnetic fluids (plasmas) was studied initially by, e.g. \cite{kruskalschwartzschild}). Their investigation showed that the magnetic field provides a stabilising effect to perturbations in the direction of the magnetic field, by supplying magnetic tension, which counteracts some of the destabilising effects of buoyancy. Clearly, this statement remains true as long as the equilibrium magnetic field has a horizontal component, however perturbations of the interface in a magnetic field that is perpendicular to the interface behave similarly to the hydrodynamic case, though with altered magnitude. The particular effect of inclination of wave propagation with respect to magnetic field direction was first investigated by \cite{chandrasekhar}. Perturbations of the horizontal equilibrium surface (or interface) give rise to oscillatory surface motion. The displaced fluid (essentially a fluid interchange) goes through a potential energy change, which will cause the interface to become unstable. Numerical investigations of RTI show the evolution of an ordered system into a stage where the two fluids mix up and small scale structures (or filaments) are developed.    

The problem of RTI in magnetic fluids (also called magnetic RTI, or MRT instability) received special attention in several astrophysical application. For example, the filamentary structure of the Crab Nebula could well be due to this instability, as was corroborated by comparisons made by \cite{hester} between the observed structuring and MRT instability simulations made by \cite{jun}. Other examples where MRT instabilities may develop are in supernovae, (e.g. \cite{fryxwell}, \cite{junnorman}, \cite{chevalier}), accretion discs, (e.g. \cite{wang} and \cite{kulkarni}), buoyant magnetic bubbles in galaxies, (e.g. \cite{robinson} and \cite{jones}) for (2D) simulations, and \cite{oneill} for (3D) and relativistic jets, e.g. \cite{M&M}.  

Parallel with the increase of high-resolution observational capabilities, MRT also received special attention in solar and solar-terrestrial physics. \cite{isobe05} and \cite{isobe06} suggested that MRT instabilities are responsible for the filamentary structure in emerging flux regions. Most notably, solar prominences are likely to become Rayleigh-Taylor unstable since they are composed of cool and dense plasma, suspended above much lighter coronal plasma. The formation of filamentary threads of prominences was modelled analytically by e.g. \cite{terradas}, who found that the RTI caused filaments in quiescent prominences to have very short lifetimes, but that in active regions, the magnetic tension may be sufficient to stabilise these prominences. MRT instabilities in prominences have been invoked by \cite{ryutova} to explain the formation of plumes and even to determine the magnetic field strength from the wavelength and growth rate of instabilities. This was expanded upon by \cite{innes}, who used the critical wavelength at which waves become unstable to estimate the Alfv\'{e}n speed. \cite{carlyle}, used the most unstable mode analysis to find the magnetic field strength in fragmenting eruptions of filaments. More recently several studies have been devoted to the investigation of the combined effects of Kelvin-Helmholtz instability upon Rayleigh-Taylor ``fingers''. This has in particular been applied to observations of plumes or bubbles within prominences (\cite{berger10}, \cite{berger17}, \cite{mishra18}, \cite{mishra19}).  For a more comprehensive review of RTI within prominences, see  \cite{hillier}). 

Over the years, many advances have been made in the theory of MRT instabilities. Compressibility effects have been widely studied in different context by (e.g. \cite{vandervoort}, \cite{shivamoggi}, \cite{bernstein},  \cite{ribeyre}, \cite{livescu}) and these studies showed univoquelly that compressibility is able to stabilise the interface against MRT instability. More recently several studies focussed on the effect of non-parallel magnetic field, that is magnetic shear at a tangential discontinuity (e.g. \cite{ruderman+t+b} and  \cite{ruderman17}). The growth time derived by these authors was found to be dependent on the shear angle and the instability increment obtained by these authors allowed them to determine that the optimal shear angle of the magnetic field was 13 degrees, based on Hinode observations see e.g. \citep{okamoto}. Since solar prominences have often temperatures that are not high enough for a complete ionisation of the plasma, the effect of partial ionisation on the generation and evolution of MRT instabilities was also studied by (e.g. \cite{diaz}, \cite{ruderman+b+k}). These investigations showed that the MRT instability becomes sensitive to the degree of plasma ionisation only for plasmas with small values of plasma beta and in a very weakly ionised state. Moreover, perturbations are unstable only for those wavenumbers that are below a cut-off value.

While these investigations into RTI are far reaching, almost all theoretical investigations are concerned with a tangential discontinuity, where the magnetic field is parallel to the interface. However, in many cases in the solar atmosphere the magnetic field is not strictly parallel to separating interfaces (e.g. sunpots' penumbra, solar prominences, plumes/interplumes in coronal holes, etc.), meaning that a new approach is needed to analyse the generation of instabilities at these interfaces. In the theory of discontinuities, we deal with tangential discontinuities when the magnetic field is parallel to the separating interface. In this case the normal component of velocity and total pressure (kinetic and magnetic) are conserved quantities across the interface. Density, kinetic pressure and tangential component of the magnetic field can be discontinuous across the interface. In contrast, when the magnetic field permeates the interface we are dealing with contact discontinuities, where the kinetic pressure, the magnetic field and velocity are all continuous quantities, while only mass density and temperature are allowed to have a jump. In an incompressible plasma, the only allowable discontinuous quantity is the mass density.  

The aim of the present work is to expand the theory of MRT instability to the case of \emph{contact discontinuities}. This has applications to arcade-type prominences, with typical ``dipped'' magnetic field structure, which intersects the denser upper plasma and so contact discontinuities become pertinent. We analytically calculate the growth rate in terms of equilibrium parameters, for the case of a contact discontinuity of an incompressible plasma.

The structure of this paper is as follows. Section 2 formulates the problem and sets out the equilibrium configuration explored in this paper. Section 3 is dedicated to the analytical derivation of the dispersion relation by solving the governing equations on both sides of the interface and then connecting the solutions using specific boundary (jump) conditions. Our analytical results are further investigated numerically in Section 4 for a number of parameters and we determine the variation of the instability growth rate with respect to the wavenumber, inclination angle of the magnetic field, propagation angle, density ratio and magnetic field strength. In Section 5 we apply our theoretical results to diagnose the inclination angle of the magnetic field based on high-resolution observations of prominence threads. Finally, we conclude and summarise our results in Section 6.


\section{Problem formulation}

The properties of MRT instability are investigated by assuming an interface in the $(x,y)$-plane situated at $z=0$, separating two distinct plasma regions, with density changing sharply at this interface according to 
\begin{equation}
\rho_0(z) = \begin{cases} \rho_1(z), & z<0, \\ \rho_2(z), & z>0. \end{cases}
\end{equation}
The contact discontinuity is attained by considering an equilibrium magnetic field that intersects the density interface at an angle, $\theta$, so we take the equilibrium magnetic field to be $\mathbf{B}_0=B_0(\cos\theta, 0, \sin\theta)$. Due to the constraint of continuity of magnetic field strength across the interface, the background magnetic field has the same value in both plasma regions and $B_0$ is constant. The equilibrium configuration is shown in Fig. 1.

The dynamics of the plasma is described using the incompressible, linearised and ideal set of MHD equations
\begin{equation}
\rho_0\pderiv{\mathbf{v}}{t}=-\nabla p +(\nabla \times \mathbf{b})\times\frac{\mathbf{B}_0}{\mu_0}-\rho g \hat{\mathbf{z}},
\label{eq:2.1}
\end{equation}
\begin{equation}
\pderiv{b}{t}=(\mathbf{B}_0\cdot\nabla)\mathbf{v},
\label{eq:2.2}
\end{equation}
\begin{equation}
\nabla\cdot\mathbf{v}=0, \quad \nabla\cdot\mathbf{b}=0,
\label{eq:2.3}
\end{equation}
where all quantities with subscript $0$ denote the equilibrium state, ${\bf b}=(b_x,b_y,b_z)$ is the perturbation of the magnetic field, ${\bf v}=(v_x,v_y,v_z)$ is the velocity perturbation, $p$ and $\rho$ are the perturbations of kinetic pressure and density, $\mu_0$ is the permeability of free space, $g=274$ m s$^{-2}$ is the constant gravitational acceleration and $\hat{\mathbf{z}}$ is the unit vector along the $z$ axis. Eq. (\ref{eq:2.3}) describes the incompressible limit and the solenoidal condition, respectively.

\begin{figure}
\centering
\includegraphics[width=0.45\textwidth]{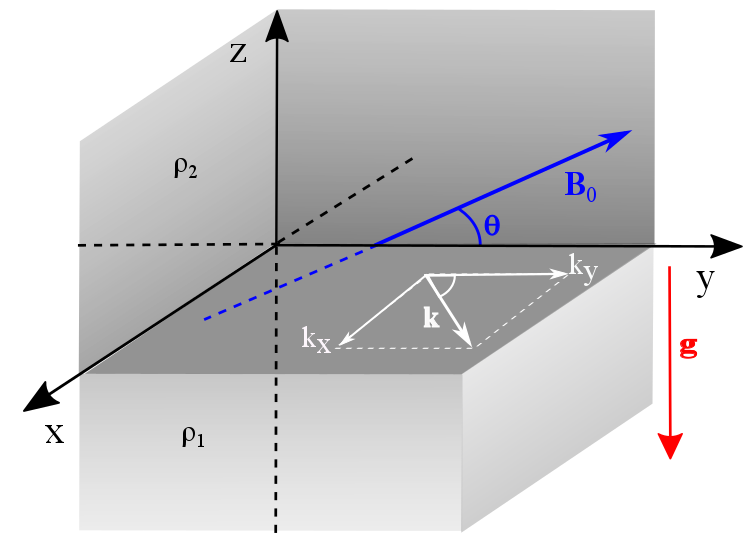}
\caption{Sketch of the equilibrium configuration used in the present work. The equilibrium state consists of a surface separating two regions, each with different density. The magnetic field is uniform throughout both regions and is inclined within the $(x,z)$-plane at an angle, $\theta$, with respect to the $x$-direction. The configuration is invariant in the $x$- and $y$-directions. Perturbations are described by the wavevector, $\mathbf{k}$, in the $(x, y)$-plane. }
\end{figure}

We consider wave-like perturbations directed along the interface situated at $z=h(x,y,t)$ and write all perturbed variables as $f\sim \hat{f}\exp[i(k_xx+k_yy-\omega t)]$. Since gravity renders the plasma to be inhomogeneous, we take the background pressure and density to depend on height alone, i.e. $\rho_0=\rho_0(z), \ p_0=p_0(z)$. Because the equilibrium magnetic field crosses the density interface, the surface $h(x,y,t)$ becomes a contact discontinuity. In order to connect the solutions of both sides across the discontinuity, we have to impose boundary conditions that are different from the ones used in the case of tangential discontinuity. In our case, first of all, we require the continuity across the interface of all three components of both velocity and the magnetic field. In the linear approximation they are given by 
\begin{equation}
\diff{\mathbf{v}}=0, \quad
\diff{\mathbf{b}}=0,
\label{eq:conts}
\end{equation}
where $\diff{{\cdot}}$ denotes the jump of a quantity across the discontinuity. 
To determine the final continuity condition let us recall that the equilibrium pressure balance is valid, i.e.
\[\deriv{p_0}{z}=-g\rho_0.\]
Assuming small changes in pressure (linear approximation) we can expand the pressure perturbation at the interface in Taylor series as
\begin{equation}
p(z=h) = p(0)+h\deriv{p_0}{z}\big|_{z=0} = p(0) - hg\rho_0(0).
\end{equation}
Hence, by differentiating, with respect to time, the condition for continuity of pressure at $z=h(x,y,t)$, we arrive at
\begin{equation}
\diff{g\rho_0v_z-\pderiv{p}{t}}= \diff{i\omega p + g\rho_0v_z}=0.
\label{eq:contp}
\end{equation}
In what follows, we solve the system of MHD equations on both sides of the interface and connect the solutions at the interface using the boundary (jump) conditions. The resulting equation will constitute the dispersion relation that we will investigate later.


\section{Dispersion relation of waves propagating at the density interface }

Considering all perturbations proportional to the exponential ansatz introduced in Section 2, Eqs. (\ref{eq:2.1}--\ref{eq:2.3}) can be written as
\begin{equation}
-i\omega\rho_0v_x = -ik_xp+\frac{B_0}{\mu}\sin\theta\left( \pderiv{b_x}{z}-ik_xb_z \right),
\label{eq:3d_momx}
\end{equation}
\[
-i\omega\rho_0v_y = -ik_yp+\frac{B_0}{\mu}\sin\theta\left( \pderiv{b_y}{z}-ik_yb_z \right)+
\]
\begin{equation}
+\frac{B_0}{\mu}\cos\theta\left( ik_xb_y-ik_yb_x\right),
\label{eq:3d_momy}
\end{equation}
\begin{equation}
-i\omega\rho_0v_z = -\pderiv{p}{z}+\frac{B_0}{\mu}\cos\theta\left( ik_xb_z-\pderiv{b_x}{z} \right) - \rho g,
\label{eq:3d_momz}
\end{equation}
\begin{equation}
-i\omega \mathbf{b} = B_0\left(ik_x\cos\theta\mathbf{v} + \sin\theta\pderiv{\mathbf{v}}{z} \right)
\label{eq:3d_induc}
\end{equation}

These equations can be combined into two relations that connect $v_x$ and $v_z$ as,
\[
i\frac{k^2}{k_x}f(v_x) =-f\left(\pderiv{v_z}{z}\right)
\]
\begin{equation}
\frac{1}{k_x}\left[f \left( \pderiv{v_x}{z} \right) + \omega^2 \frac{\rho_0'}{\rho_0}v_x \right] =i\left[f(v_z) + g \frac{\rho_0'}{\rho_0}v_z\right],
\label{eq:3D_simpl_vx_vz}
\end{equation}
where the operator function $f$ is defined as
\begin{equation}
f = v_A^2\sin^2\theta \pderiv{^2}{z^2} + ik_xv_A^2\sin2\theta\pderiv{}{z} + (\omega^2-k_x^2v_A^2\cos^2\theta).
\end{equation}
Using Eq. (\ref{eq:3d_momx}), we can now express the kinetic pressure perturbation as
\[
p =\left[\frac{1}{k_x}\frac{\rho_0v_A^2}{\omega}\sin^2\theta\pderiv{^2}{z^2} + i\frac{\rho_0v_A^2}{2\omega}\sin2\theta\pderiv{}{z}+\frac{\omega}{k_x}\rho_0\right]v_x+
\]
\begin{equation}
+\rho_0v_A^2\left[-\frac{i}{\omega}\sin^2\theta\pderiv{}{z} + \frac{k_x}{2\omega}\sin2\theta\right]v_z.
\label{eq:3d_p}
\end{equation}

When the vertical scales of perturbations are much smaller than the gravitational scale-height, such as at the onset of instability in the given problem, we can take the long scale-height limit and so treat the plasma either side of the interface as uniform. Given this approximation, we will simplify our treatment by considering $\rho_0'=0$, therefore we can combine the equations given by (\ref{eq:3D_simpl_vx_vz}), into the \emph{governing equation} of wave propagation in the incompressible plasma, 
\begin{equation}
\left(k^2-\pderiv{^2}{z^2}\right) f(v_z) =0.
\label{eq:3d_gov_1}
\end{equation}
We assume that the solution of the above equation is of the form $\hat{v}_z \sim e^{\Gamma z}$, where the quantity $\Gamma$ is a complex quantity. Eq. (\ref{eq:3d_gov_1}) then simplifies to
\begin{equation}
(\Gamma^2-k^2)(v_A^2\sin^2\theta \ \Gamma^2 + ik_xv_A^2\sin2\theta  \Gamma + \omega^2 - k_x^2v_A^2\cos^2\theta )=0,
\label{eq:3d_gov}
\end{equation}
which admits the roots
\[\Gamma = \pm k, m_{\pm},\]
where the expression of $m_\pm$ is given by
\begin{equation}
m_\pm = \frac{-i\left[v_Ak_x\cos\theta\pm\omega\right]}{v_A\sin\theta}.
\end{equation}

The form of the governing equation derived in Eqs. (\ref{eq:3D_simpl_vx_vz}) imply that the frequency of waves will be complex, with the imaginary part of $\omega$ describing the temporal evolution of perturbations' amplitude. According to the temporal variation of perturbations assumed earlier, it is clear that a positive imaginary part of $\omega$ will describe unstable amplification of the amplitude. Since we are interested in waves being localised at the surface, we assume that far away from the interface waves will be evanescent, therefore, the $z$-component of the velocity will be of the form
\begin{equation}
\hat{v}_z = \begin{cases} A_1 \ e^{kz} + B_1 \ e^{m_{+1}z},\  z<0, \\ A_2 \ e^{-kz} + B_2 \ e^{m_{-2}z}, \ z>0, \end{cases}
\label{eq:3d_vz_exp}
\end{equation}
where $A_1$, $A_2$, $B_1$ and $B_2$ are arbitrary constants.  Below we are only interested in unstable perturbations. Hence we assume that $\Im(\omega)>0$. This will also ensure that waves will be evanescent (via the roots $m_\pm$). While the first terms in Eq. (\ref{eq:3d_vz_exp}) represent an exponential decay with $z$, the second terms describe an oscillatory decay. It is also obvious that when $\theta= 0$, the second terms of (\ref{eq:3d_vz_exp}) vanish and we recover the result of a tangential discontinuity.

The present analysis, where the two plasma regions are treated as homogeneous, is valid when $1/|\Re(\Gamma)|$, is much less than the gravitational scale-height, $H$, i.e. 
\[
H\gg \frac{1}{k}, \quad \frac{v_{A1}\sin\theta}{\omega_i}, \quad \frac{v_{A2}\sin\theta}{\omega_i}.
\] 
In other words, stratification effects may be ignored, provided the wavelength of perturbations as well as the product of the instability time with the Alfv\'en speeds are sufficiently small.

With the help of the expressions (\ref{eq:3d_vz_exp}) we can find the corresponding equations for the pressure and magnetic field perturbations. The solutions obtained for the two regions will be connected at the interface using the boundary conditions specified earlier.

\subsection{Boundary conditions}

Using the surface wave solutions (\ref{eq:3d_vz_exp}), we can employ the boundary conditions given by Eqs. (\ref{eq:conts}) and (\ref{eq:contp}) to find the dispersion relation of incompressible waves propagating along the interface. 
We may now express these conditions (including Eq. \ref{eq:contp}) in terms of $\hat{v}_x$ and $\hat{v}_z$.

The incompressibility condition yields
\[
\frac{\partial v_z}{\partial z}=-i(k_xv_x+k_yv_y).
\]
Taking into account the behaviour of $v_x$ and $v_y$ at the interface, we can obtain that, in addition, that the condition
\begin{equation}
\diff{\pderiv{v_z}{z}}=0
\label{eq:3d_cont_dvz}
\end{equation}
has to be also satisfied. Imposing the continuity of $b_x$ across the interface together with Eq. (\ref{eq:3d_induc}) results in
\[ \diff{i\sin\theta\pderiv{v_x}{z}-k_x\cos\theta \ v_x}=0.\] 
Combining this condition with continuity of $v_x$, we obtain
\begin{equation}
\diff{\pderiv{v_x}{z}}=0.
\label{eq:3d_cont_dvx}
\end{equation}
Furthermore, based on the same consideration, the boundary condition written for $b_y$ simplifies to
\begin{equation}
\diff{\pderiv{^2v_z}{z^2}}=0.
\label{eq:3d_cont_d2vz}
\end{equation}
Finally we can use the expression we derived for the kinetic pressure (Eq. \ref{eq:3d_p}) together with the jump conditions derived above to transform the dynamic boundary condition (\ref{eq:contp}) as
\begin{equation}
\diff{i\rho_0\left(v_A^2\sin^2\theta\pderiv{^2v_x}{z^2} + \omega^2v_x\right)+k_xg\rho_0v_z}=0.
\label{eq:3d_cont_p}
\end{equation}
The above boundary conditions will be used to connect the solutions from both sides of the interface in order to derive the dispersion relation of waves. 

\subsection{Derivation of Dispersion Relation}

In order to use the boundary conditions specified above, we must first find the expression for $v_x$. Using the expression of $v_z$ given by Eq. (\ref{eq:3d_vz_exp}), we can solve the differential equation for $v_x$ (see the first equation of the system \ref{eq:3D_simpl_vx_vz}), and we obtain that
\begin{align}
\hat{v}_x = \begin{cases} i\frac{k_x}{k}A_1 \ e^{kz} + C_1\ e^{m_{+1}z} & z<0, \\ -i\frac{k_x}{k}A_2 \ e^{-kz} + C_2 \ e^{m_{-2}z} & z>0. \end{cases}
\label{eq:3d_vx_exp}
\end{align}
where $C_1$ and $C_2$ are more arbitrary constants.

Using the exponential forms of $v_x$ and $v_z$ specified by Eqs. (\ref{eq:3d_vz_exp}) and (\ref{eq:3d_vx_exp}), we can now apply the boundary conditions (Eq. (\ref{eq:3d_cont_dvz})-- (\ref{eq:3d_cont_p}) along with $\diff{v_x}=0$ and $\diff{v_z}=0$). After long but straightforward calculations we obtain a system of linear equations for the unknown six constants, that in matrix form, can be given as
\begin{equation}
\begin{bmatrix}
1 & 1 & 0 &-1 &-1 &0 \\
i\frac{k_x}{k}& 0 & 1 & i\frac{k_x}{k}& 0 & -1\\
k & m_{1+} & 0 & k & -m_{2-} & 0\\
k^2 & m_{1+}^2 & 0 & -k^2 & -m_{2-}^2 & 0\\
ik_x & 0 &m_{1+} & -ik_x &0 &-m_{2-}\\
\alpha_1 & \beta_1 & \gamma_1 & -\alpha_2 & -\beta_2 & -\gamma_2
\end{bmatrix}
\begin{bmatrix}
A_1 \\ B_1 \\ C_1 \\ A_2 \\ B_2 \\C_2
\end{bmatrix} = M\begin{bmatrix}
A_1 \\ B_1 \\ C_1 \\ A_2 \\ B_2 \\C_2
\end{bmatrix} =0,
\label{eq:matrix}
\end{equation}
where the various entries of the matrix $M$ are defined as
\[
\alpha_1 = k_xg\rho_1-\rho_1\frac{k_x}{k}\left(v_{A1}^2k^2\sin^2\theta +\omega^2\right),
\]
\[
\alpha_2 = k_xg\rho_2+\rho_2\frac{k_x}{k}\left(v_{A2}^2k^2\sin^2\theta +\omega^2\right),
\]
\[
\beta_1 = k_xg\rho_1, \quad \beta_2 =  k_xg\rho_2,
\]
\[
\gamma_1 = i\rho_1\left(v_{A1}^2m_{1+}^2\sin^2\theta +\omega^2\right),  \quad \gamma_2 =  i\rho_2\left(v_{A2}^2m_{2-}^2\sin^2\theta +\omega^2\right),
\]
and $k=(k_x^2+k_y^2)^{1/2}$. The non-trivial solution to the homogeneous system of equations (\ref{eq:matrix}) only exists when $\det(M)=0$, which will give us the dispersion relation of wave perturbations propagating along the interface in the presence of gravity and inclined magnetic field as
\begin{equation}
(m_{2-}-m_{1+})(k-m_{1+})(k+m_{2-})S=0,
\label{eq:disprel}
\end{equation}
where
\begin{align}
S = &(d^{1/2}+1)(d-1)\omega^3
-2ikd^{1/2}v_{A1}\sin\theta(d^{1/2}+1)^2\omega^2\nonumber\\
&+(d^{1/2}+1)[2dv_{A1}^2(k^2-k_y^2\cos^2\theta) + gk(d-1)]\omega\nonumber\\
&+2ik^2d^{1/2}v_{A1}g\sin\theta(d-1),
\label{eq:S}
\end{align}
and $d=\rho _1/\rho_2<1$ is the density ratio. The dispersion relation (\ref{eq:disprel}) has five non-trivial solutions (the first multiplier in the dispersion relation corresponds to a trivial solution) in terms of $\omega$. The second and third brackets admit the two solutions,
\[
\omega_1 = k_xv_{A1}\cos\theta - ik v_{A1}\sin\theta
\]
\begin{equation}
\omega_2 = k_x\sqrt{d}v_{A1}\cos\theta - ik \sqrt{d}v_{A1}\sin\theta.
\label{eq:3d_S1}
\end{equation}
In order to simplify the discussion, let us introduce the propagation angle within the $(x,y)$-plane, $\alpha$, such that  
\[
k_x = k\cos\alpha, \quad k_y = k\sin\alpha.
\]
In addition, we use one single Alfv\'en speed ($v_{A1}$), and write the Alfv\'en speed in region 2 as, $v_{A2}=\sqrt{d}v_{A1}$. We

Clearly, $\omega_1$ and $\omega_2$ have negative imaginary parts and these solutions are disregarded, because we only consider $\Im(\omega)>0$. 
Hence, we consider the three solutions to $S=0$. In order to gain more information about the unstable solutions, this may be reformatted in terms of $\Omega=-i\omega$, such that $\Re(\Omega)=\Im(\omega)$ and the unstable solutions are given by solutions where the real part of $\Omega$ is positive. We find that $S=-i\sigma$, where
\begin{align}
\sigma =&(d^{1/2}+1)(1-d)\Omega^3
+2kd^{1/2}v_{A1}\sin\theta(d^{1/2}+1)^2\Omega^2\nonumber\\
&+(d^{1/2}+1)[2dv_{A1}^2(k^2-k_y^2\cos^2\theta) - gk(1-d)]\Omega\nonumber\\
&-2k^2d^{1/2}v_{A1}g\sin\theta(1-d).
\end{align}
Given that the coefficients of $\Omega^3$ and $\Omega^2$ are both positive (when $d<1$), the sum of the roots must be negative and since the coefficient of $\Omega^0$ is negative, the product of the roots must be positive. Assuming that at least one solution is unstable, i.e. positive $\Omega$, these two conditions lead to the fact that the other two roots for $\Omega$ must have negative real parts and so are non-physical, amplifying modes. This leaves one physical solution, that we display graphically, to explore the effects of varying wavenumber, $k$, inclination angle, $\theta$, density ratio, $d$, and propagation direction, $\alpha$, shown in Figures (\ref{fig:3d_k_dep} - \ref{fig:contour_alpha_vA}).


\section{Solutions and results}

\begin{figure}
\centering
\includegraphics[width=.5\textwidth]{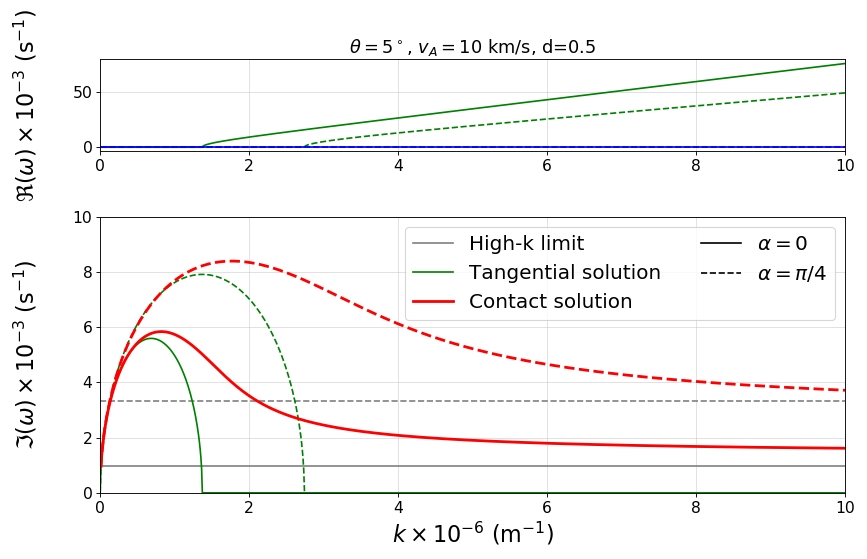}
\caption{Solutions for the dispersion relation for waves propagating in the $(x,y)$-plane, for two propagation directions ($\alpha=0$ solid line, and $\alpha=\pi/4$ dashed line), with respect to wavenumber, $k$.  The density ratio is $d=0.5$, the Alfv\'{e}n speed of the lower plasma is $v_A=10$ km s$^{-1}$, and the magnetic field inclination angle $\theta = 5^{\circ}$. The upper panel shows the real part of the frequency, while the imaginary part is plotted in the lower panel. The limiting value of $\Im(\omega)$ when $k\to \infty$ is shown by the grey horizontal lines, in the lower panel. For illustration we also show the real and imaginary part of the frequency in the case of a tangential discontinuity ($\theta=0$), plotted here in green.}
\label{fig:3d_k_dep}
\end{figure}

The variation of the only physically acceptable root of $S(\omega)=0$ with various physical parameters is investigated numerically. In Figure (\ref{fig:3d_k_dep}), we plot the variation of the real and imaginary part of the frequency with respect to the wavenumber, $k$, for a small field inclination angle. In order to compare our results with the well-known results obtained in the case of a tangential discontinuity, we plot the results we obtain for $\theta=0$ (green lines). We choose two values for the propagation direction: $\alpha=0$ (propagation along the $x$-axis, solid line) and $\alpha=\pi/4$ (dashed line). The smallest instability increment (longest amplification time) is obtained for propagation parallel to the $x$-axis, while increasing the direction of propagation away from $\alpha=0$ we see an increase in the instability rate. The maximum of the instability rate is obtained at a smaller wavelength than in the case of a tangential discontinuity, and in general, the maximum of the rate is higher than the one obtained for tangential discontinuity. This result is easy to interpret, as at contact discontinuity only the horizontal component of the equilibrium magnetic field tends to stabilise the plasma.

In the case of a tangential discontinuity, there is always a critical wavenumber, below which the solution is purely imaginary and hence gives rise to instability. However, above this critical value, the solution is real and the wave is propagating. In contrast, for the contact discontinuity, this is no longer the case; solutions are unstable for all values of wavenumber, $k$. Moreover, for very large wavenumbers the imaginary part of the frequency tends towards a fixed value given by
\begin{equation}
\omega \to \frac{ig\sin\theta(1-d)}{v_{A1}d^{1/2}(1-\sin^2\alpha\cos^2\theta)(d^{1/2}+1)}.
\label{eq:high_k}
\end{equation}
These values are shown in grey in Fig. (\ref{fig:3d_k_dep}).  It may also be seen that the instability rate decreases for higher wavenumber.

Interestingly, the behaviour of the imaginary part of the frequency (and the disappearance of the critical wavenumber) is similar to the results obtained by  \cite{diaz}, where these authors studied the effect of partial ionization on MRT instability in a single-fluid approximation. In their study, the change in the imaginary part of the frequency was attributed to the ambipolar diffusion in the induction equation, i.e. the modification was due to the presence of neutrals that can diffuse in the perpendicular direction to the ambient magnetic field. This suggests that changes occurring in the transversal direction (relative to the interface) will notably modify the behaviour of the instability increment. One important difference is that, for the partially ionised case considered by \cite{diaz}, the instability rate tends to zero for high wavenumbers, however, in our case this quantity never reaches the zero value. 

\begin{figure}
        \centering
        \includegraphics[width=.5\textwidth]{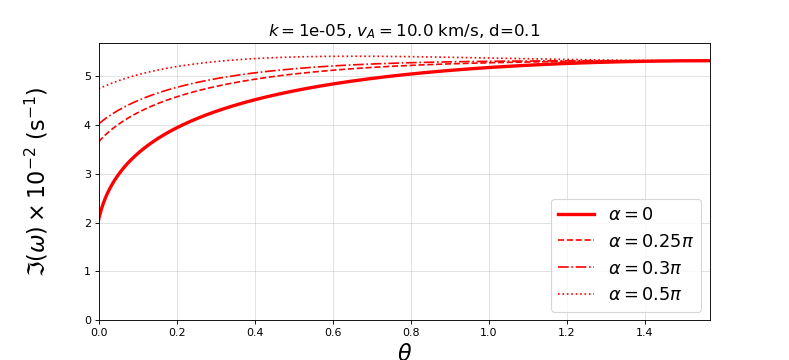}
        \caption{Imaginary part of the frequency for waves propagating in the $(x,y)$-plane, for several propagation directions, with respect to the magnetic field inclination, $\theta$.  The density ratio is assumed to be $d=0.5$, the reference Alfv\'{e}n speed is $v_{A1}=10$ km s$^{-1}$, and the value of the wavenumber fixed at $k= 10^{-5}$ m$^{-1}$.}
\label{fig:3d_theta_dep}
\end{figure}

In the other limit, when the propagation direction is along the $y$-axis ($\alpha=\pi/2$), the only wave that can propagate is the gravity surface wave. The magnetic field begins to have an effect on the propagation characteristics of waves, as the propagation direction inclines towards the $x$-axis.

Let us investigate how the inclination angle of the magnetic field, $\theta$, affects the stability of incompressible waves propagating along the interface. Now, the value of the wavenumber is fixed at $k= 10^{-5}$ m$^{-1}$ (a typical value for oscillations in prominences) and we choose a number of characteristic values for the propagation angle, $\alpha$. The numerical result of our analysis is shown in Fig. (\ref{fig:3d_theta_dep}). Again, the mode that has the smallest instability rate is the one that propagates strictly along the $x$-axis and this rate shows a pronounced increase for smaller values of $\theta$, after which this rates saturates and becomes independent of the inclination angle of the magnetic field. With increasing the propagation angle the instability rate increases, meaning that the amplification time reduces. This result is something that can be understood if we keep in mind that, with increasing the propagation angle, the magnetic tension has less effect on the stabilisation of the interface. It is clear that regardless of the propagation angle of waves, the instability increments tend to a steady value of approximately $\Im(\omega)=0.055$ s$^{-1}$.

\begin{figure}
    \centering
        \includegraphics[width=.5\textwidth]{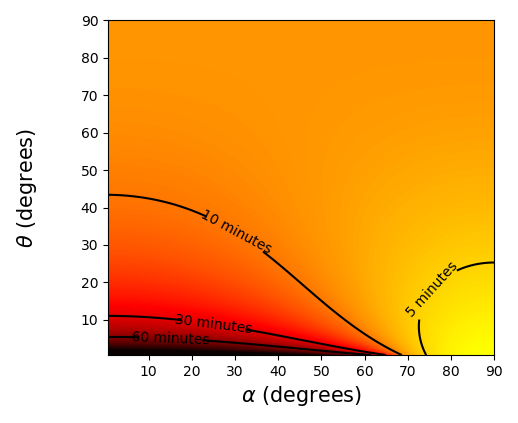}
        \caption{Growth time for $k = 10^{-5}$ m$^{-1}$, $d=0.5$ $v_A=10$ kms$^{-1}$, in terms of propagation angle, $\alpha$ and field inclination $\theta$.}
        \label{fig:contour_theta_alpha}
\end{figure}

In Fig. (\ref{fig:3d_theta_dep}), we see that regardless of the value of the propagation angle of the waves, for large inclination angle of the field, all modes will tend towards the same instability rate. 
In order to translate our results into observable quantities, we show the contour plots of the inverse of the instability rate (the growth time, in minutes) for a particular wavenumber, density ratio, $d$ and Alfv\'en speed in terms of the inclination angle, $\theta$ and propagation angle, $\alpha$ (see Fig. \ref{fig:contour_theta_alpha}), i.e. we plot the pair of the angles that satisfy the given growth time. In this plot, we see two distinct behaviours. While on the left-hand side of the plot we see that for shorter growth time we require higher inclination angle, the mode that appears on the right-hand side shows a different behaviour for very large values of inclination angle. However, this mode is not an Alfv\'en mode; instead it is the surface gravity mode that appears for a nearly perpendicular propagation. Similarly to the findings shown in Fig.(\ref{fig:3d_theta_dep}), for large inclination angle of the magnetic field, the growth time of instability becomes independent of $\theta$.

\begin{figure}
    \centering
       \includegraphics[width=.5\textwidth]{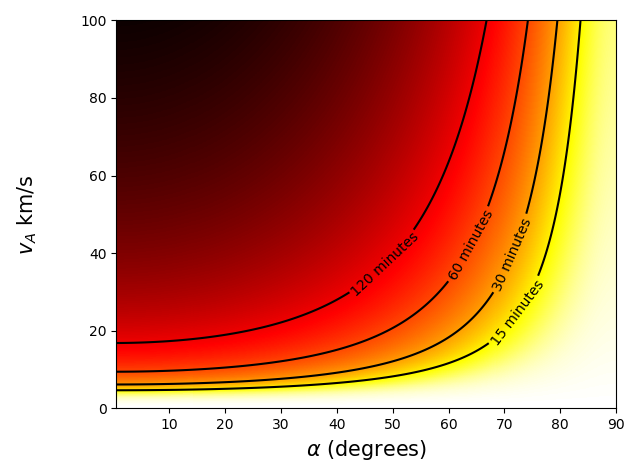}
        \caption{Growth time for $k = 10^{-5}$ m$^{-1}$, $d=0.5$ $\theta=5^{\circ}$, in terms of propagation angle, $\alpha$ and Alfv\'{e}n speed of the lower plasma, $v_A$.}
        \label{fig:contour_alpha_vA}
\end{figure}

It is well known that magnetic field can stabilise the unstable interface, that is why we investigate the variation of the growth time with respect to the propagation angle of waves and Alfv\'{e}n speed, keeping the wavenumber, density ratio and magnetic field inclination constant (see Fig.(\ref{fig:contour_alpha_vA})). Since the density ratio is constant, changing Alfv\'{e}n speed means a change in the intensity of the magnetic field. An increase in the magnetic field intensity means that the growth time increases, i.e. the magnetic field has a stabilising effect, for any propagation direction, as expected. In addition, the value of the propagation angle becomes important for stronger magnetic fields.

Finally, we investigate the effect of changing density ratio on the growth rate of unstable modes and we obtain that the growth rate decreases monotonically towards zero as the density ratio is increased, however, maintaining the $d<1$ condition. In the case of the tangential MRT instability (but also true for the hydrodynamic case), as the density of the upper plasma increases in comparison to the density of the lower plasma ($d$ decreasing), the system becomes less stable. Taking the limit of $d \to 0$, the instability rate tends towards a fixed value. By taking the solution of $S(\omega)=0$, when $d \to 0$, the limiting value of the growth rate is found to be $\Im(\omega) \to \sqrt{gk}$. This is clearly independent of both inclination angle and magnetic field strength.


\section{Applications to Solar Prominences}

Solar prominences are magnetic features suspended in the solar corona that are made from dense, cold plasma surrounded by tenuous and hot coronal plasma. High-resolution observations show that prominences present threads along which plasma can flow and waves propagate.  Such an observation was carried out, using Hinode/SOT instrument by \cite{okamoto}, who observed  an active region prominence (NOAA AR 10921) in a 0.3 nm broadband region centred at 396.8 nm. They found that a multi-thread prominence was suspended above the main sunspot. Their analysis showed in-phase oscillatory motion with periods 130-250 s. These authors concludes that the observed waves are propagating or standing Alfv\'en waves. 

\cite{terradas08} used these oscillations to carry out a seismological study and determined Alf\'en speeds in both the prominence and corona in terms of the density ratio. The threads were considered to be thin flux tubes (observations showed that they have a small radius) in the presence of a flow that proved to slightly influence the period of waves. Using a few simplifications (e.g. straight flux tubes, homogeneous plasma, longitudinal magnetic field, non-stratified plasma, threads have equal length, linear approximation) these authors obtained that the minimum Alfv\'en speed varies between 120 and 350 km s$^{-1}$. Although not analysed in the above studies, \cite{ruderman+t+b}, found that the typical lifetimes of threads under investigation is 10 minutes. If the MRT is responsible for these short lifetimes, then we may suppose that the instability time is approximately equal to the thread lifetime. This gives us the necessary information to find the magnetic field inclination, $\theta$, in terms of the propagation direction $\alpha$, for a given density ratio. 

\begin{figure*}
\centering
\begin{subfigure}{0.3\textwidth}
\includegraphics[width=\textwidth]{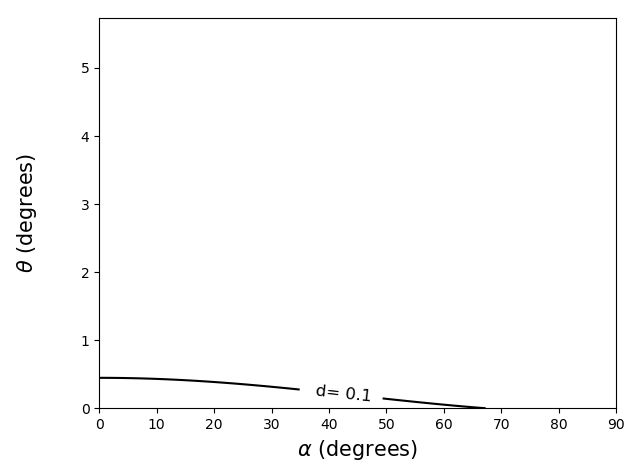}
\caption{Thread 1}
\end{subfigure}%
\begin{subfigure}{0.3\textwidth}
\includegraphics[width=\textwidth]{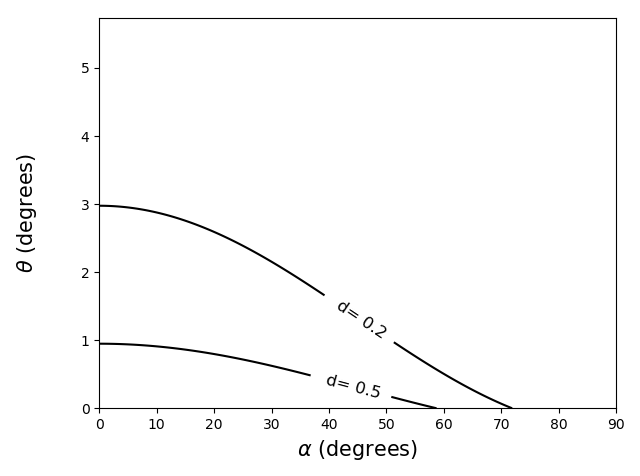}
\caption{Thread 2}
\end{subfigure}%
\begin{subfigure}{0.3\textwidth}
\includegraphics[width=\textwidth]{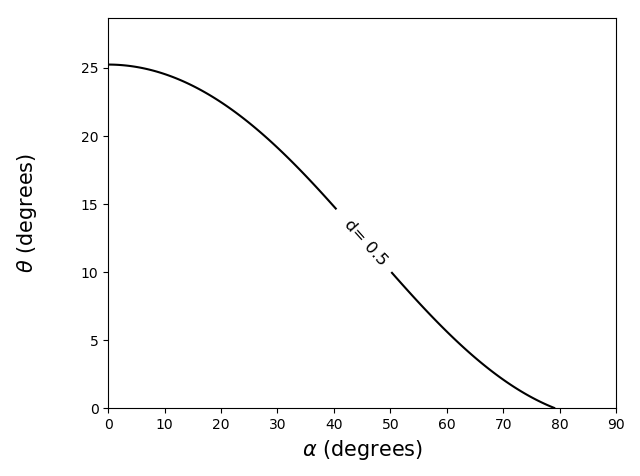}
\caption{Thread 3}
\end{subfigure}\\
\begin{subfigure}{0.3\textwidth}
\includegraphics[width=\textwidth]{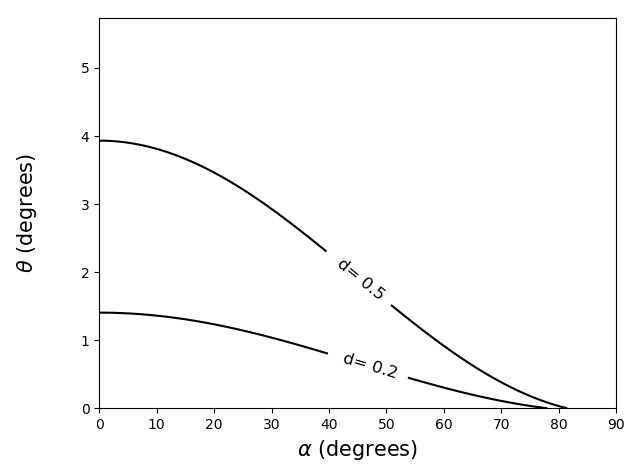}
\caption{Thread 4}
\end{subfigure}%
\begin{subfigure}{0.3\textwidth}
\includegraphics[width=\textwidth]{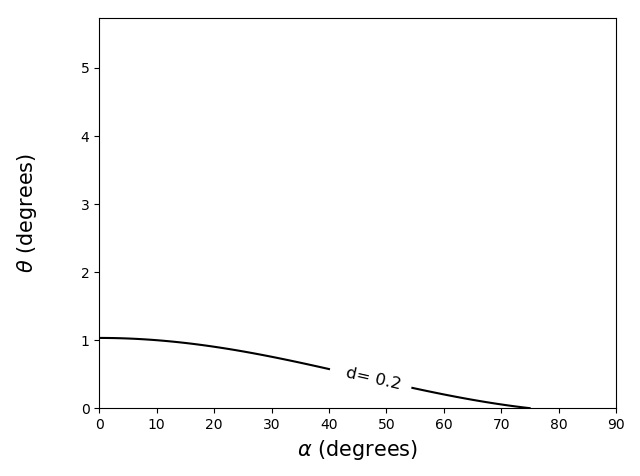}
\caption{Thread 5}
\end{subfigure}%
\begin{subfigure}{0.3\textwidth}
\includegraphics[width=\textwidth]{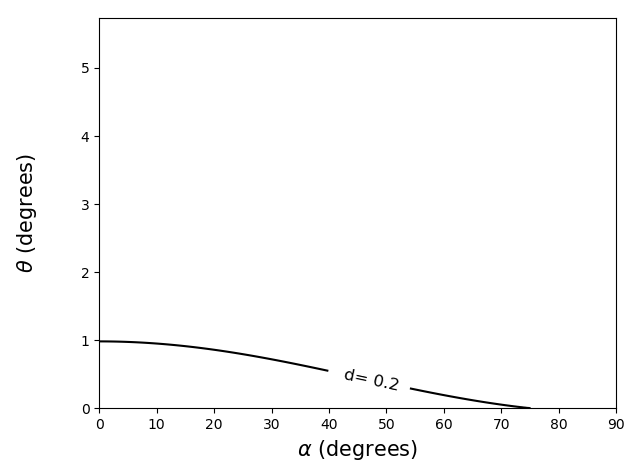}
\caption{Thread 6}
\end{subfigure}
\caption{Solutions of the dispersion relation for a given growth time in terms of magnetic field inclination ($\theta$) and propagation direction ($\alpha$), for six observed prominence threads, for three possible density ratios. Observational data has been adapted from \citep{okamoto}. }
\label{fig:prominences}
\end{figure*}

Using the observations by \cite{okamoto} we can use the analysis presented by \cite{terradas08} (their Eq. 3) to determine the wavenumbers and Alfv\'en speeds for three possible density ratios ($d=0.1, \ 0.2, \ 0.5$). With these values we solve numerically Eq. (\ref{eq:S}) and this determines the pair of values for magnetic field inclination and propagation direction satisfying the observed variables and the governing equation (\ref{eq:S}). We note that not all of observable data sets have solutions for any $\alpha$ and $\theta$ values, suggesting that only certain density ratios may be possible in those circumstances, thus giving even more information about hard-to-observe variables. The solutions, in terms of magnetic field inclination and propagation direction, are displayed for all six observed threads in Figure (\ref{fig:prominences}) and the relevant data, including the maximum possible inclination angle (when the wave is directed in the same plane as the magnetic field i.e $x$-propagating), are shown in the table.

It is clear that for all threads the 10 min growth time can be satisfied only for particular values of density ratio and particular combination of field inclination angle and propagation angle. Thread 1 shows only solutions for a density ratio of $d=0.1$, with the maximum field inclination only about half a degree that is attained when the wave propagates along the $x$-axis. Threads 4, 5 and 6 all have solutions for $d=0.2$ with maximum inclination of approximately $1^{\circ}$. In addition thread 4 has a solution for $d=0.5$ with a noticeably higher maximum inclination of $4^{\circ}$. Thread 2 shows a higher maximum $\theta$ for the $d=0.2$ solution than the other three, of $3^{\circ}$, and a lower maximum $\theta$ for the $d=0.5$ solution than thread 4 of approximately $1^{\circ}$. Finally, thread 3 has a solution only for $d=0.5$, but with much higher maximum inclination than any of the other threads, with $\theta=25^{\circ}$. Although all of the magnetic field inclinations found are relatively low (except prominence 3), the inclination is only zero if the propagation direction is almost perpendicular to the magnetic field, which would be very unlikely for MHD waves in the solar atmosphere. This also gives a justification for the study carried by \cite{vickers}, where small magnetic field inclinations were considered for MHD waves at a contact discontinuity. 

\begin{figure*}
\centering
\begin{subfigure}{0.5\textwidth}
\includegraphics[width=\textwidth]{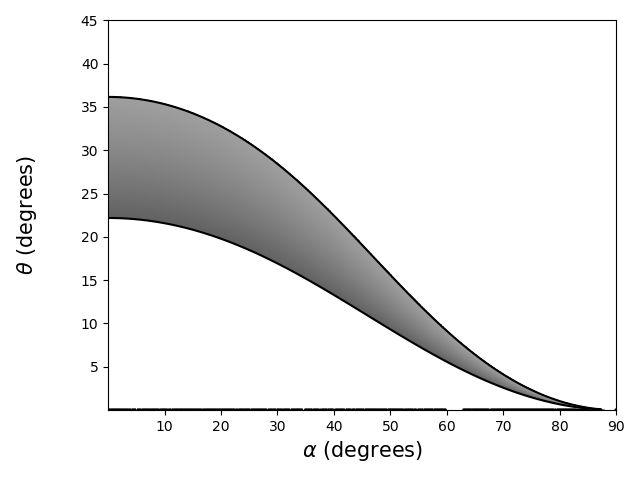}
\caption{Plume 1}
\end{subfigure}%
\begin{subfigure}{0.5\textwidth}
\includegraphics[width=\textwidth]{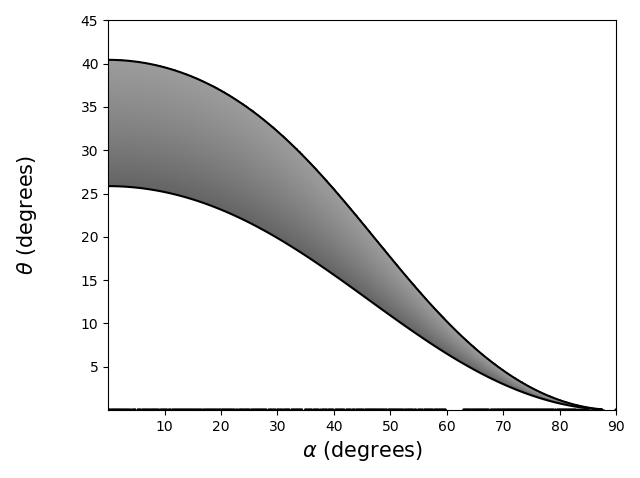}
\caption{Plume 2}
\end{subfigure}
\caption{Solutions of the dispersion relation for a given growth time in terms of magnetic field inclination ($\theta$) and propagation direction ($\alpha$), for two observed prominence plumes. Observational data has been taken from \citep{mishra19}.} 
\label{fig:plumes}
\end{figure*}

In order to expand our analysis, we also consider the case of plumes, seen to develop in prominences. In particular we look at two plumes observed by \cite{mishra19}, where growth rates of plumes were estimated to be $1.32 \ \pm \ 0.29 \ \times \ 10^{-3}$ s$^{-1}$ and $1.48 \ \pm \ 0.29 \ \times \ 10^{-3}$ s$^{-1}$. Using these values, along with the approximate wavelength, densities and Alfv\'en speed found in \cite{mishra19}, an inversion is performed using Eq. (\ref{eq:S}) and results are shown in Fig. (\ref{fig:plumes}). For this case, we find much higher  maximum inclination of $22-41^\circ$. This shows the variation of inclination possible in prominences. It also seems logical that inclination would be greater after plumes begin to develop, so our results are consistent with this assumption. 

We should note that, in the context of solar prominences, compressibility would have a pronounced effect in stabilising the interface, therefore, a model including compressibility would need to be developed to give more pertinent results. The density ratios present in prominences also, in general, give lower $d$ values than those considered here, destabilising the system further. We can thus suppose that solutions would be of a similar order of magnitude to those considered here.

\begin{table*}
\begin{center}
\begin{tabular}{|c|c|c|c|c|c|c|c|}
\hline
 & Wavenumber & \multicolumn{2}{c|}{d=0.1} & \multicolumn{2}{c|}{d=0.2} & \multicolumn{2}{c|}{d=0.5}
\\
\# & (10$^{-7}$ m$^{-1}$) & $v_A$ (km s$^{-1}$)& max $\theta$ ($^{\circ}$) & $v_A$ (km s$^{-1}$)& max $\theta$ ($^{\circ}$) & $v_A$ (km s$^{-1}$) & max $\theta$ ($^{\circ}$)\\
\hline 
1 & 34.9 & 1331 & 0.5 & 1237 & - & 1172 & - \\
2 & 7.85 & 1336 & - & 1015 & 2.9 & 897 & 1.1 \\
3 & 18.8 & 1116 & - & 983 & - & 899 & 25.8 \\
4 & 57.1 & 1220 & - & 1164 & 1.4 & 1126 & 4.0\\
5 & 35.9 & 1710 & - & 1583 & 1.1 & 1511 & -\\
6 & 7.39 & 861  & - & 827 & 1.1 & 806 & -\\
\hline
\end{tabular}
\end{center}
\label{tab:prominences}
\caption{The wavenumber and coronal Alfv\'en speeds derived for the six prominence threads. The maximum value of the inclination angle of the field is shown for the three values of the density ratio, $d$.}
\end{table*}

\section{Conclusions}

Given its importance in re-organisation of the plasma and formation of small-scale structures, the magnetic Rayleigh-Taylor instability is one of the most studied instabilities in solar and solar-terrestrial plasmas. In the present paper, we have studied the effect of an oblique magnetic field that crosses an interface in the development of incompressible MHD Rayleigh-Taylor instability. In this configuration, the sharp interface separating plasma regions with different densities forms a contact discontinuity.

We have shown that, in contrast to well-known results for tangential discontinuity, for even a small inclination of the magnetic field, the system is unstable for any wavenumber, i.e. there is no longer a ``critical wavenumber'', above which the system is stable. The instability rate is also higher than for the tangential case, however, it is lower than for the purely hydrodynamic case and is reduced with increasing Alfv\'{e}n speed, as the magnetic tension has a greater component stabilising against gravity. The instability rate was found to be considerably higher for perturbations perpendicular to the plane in which the magnetic field is inclined, than for perturbations in the direction of the field and there is a smooth change between these two extremes, as the magnetic tension has a less pronounced effect as a restoring force. 

The high density gradients at the edges of prominences, between the low density coronal plasma and the higher density plasma of the prominence itself (readily modelled, on the small scale, by a sharp interface) is often intersected by inclined magnetic fields. Therefore, we have applied the present results to model MRT instabilities in solar prominences. These results suggest that, even with a high magnetic field strength, instabilities are able to develop for perturbations of any wavelength. Thus, inversion techniques relying on the critical wavelength may not always be valid, as there is no such critical wavelength for the contact discontinuity. This also gives us information about where in the prominences plumes are most likely to develop, as contact discontinuities appear most frequently at the sides of prominences. 

Using the dispersion relation derived in the current study an inversion technique may be performed on observations of oscillating prominences, and this can give us information on the angle of inclination of the magnetic field and direction of the wave propagation. A simple and approximate analysis of six prominence threads observed by Hinode/SOT is performed and the results confirm there is a combination of the two angles that can account for the observed growth time of instabilities. In the cases studied, the typical inclination angles (where waves propagate in the direction of the plane of inclination) were found to be $1 \ - \ 4^\circ$. The inversion also provided information about possible density ratios between the prominence plasma and sparse coronal plasma, for each prominence thread. When we apply this inversion to two prominence plumes, observed by \cite{mishra19}, the average maximal field inclination angle (for perturbations directed in the plane of inclination) was found to be approximately 30$^\circ$.

This study has provided a first step in the understanding of magnetic Rayleigh-Taylor instabilities at a contact discontinuity. This study used a linear technique, which gives relevant information about the onset of instabilities, but as the instability evolves, non-linear effects will need to be taken into account, including secondary Kelvin-Helmholtz instabilities as e.g. the plumes develop. To take these into account, a time-dependent analysis would need to be performed, which, due to its complexity, must be performed numerically. We restricted our investigation to a local analysis, where the effect of gravitational stratification has been neglected, however, this approximation might not always be valid as at later stages, as e.g. plumes begin to develop, gravitational stratification will become more relevant. Compressibility is known to stabilise many plasma configurations, while partial ionisation has been shown to aid stability in certain circumstances and destabilise in others, hence both these effects should also be taken into account to obtain a more detailed insight into the effect of field inclination on MRT instabilities.

\begin{acknowledgements}
The authors acknowledge the help of M.S. Ruderman.\\
      E.V. acknowledges the support by Science and Technology Facilities Council (STFC). \\
      R.E. is grateful to Science and Technology Research Facilities Council (STFC, grant number ST/M000826/1) UK for the support received.  
\end{acknowledgements}



\begin{thebibliography}{}

\bibitem[Berger et al. (2010)]{berger10} Berger, T.E., Slater, G., Hurlburt, N. et al., 2010,
ApJ, 716, 2, 1288

\bibitem[Berger et al. (2017)]{berger17} Berger, T.E., Hillier, A., Liu, W., 2017,
ApJ, 850, 60

\bibitem[Bertstein \& Book (1983)]{bernstein} Bernstein, I. B. \& Book, D. L., 1983,
Phys. Fluids, 26, 453

\bibitem[Carlyle et al.(2014)]{carlyle} Carlyle, J., Williams, D. R., van Driel-Gesztelyi, L. et al., 2014,
ApJ 782, 87

\bibitem[Chandrasekhar (1961)]{chandrasekhar} Chandrasekhar,
1961, S. \textit{International Series of Monographs on Physics}

\bibitem[Chevalier (1982)]{chevalier} Chevalier, R. A., 1982,
ApJ, 258, 790

\bibitem[Diaz et al. (2014)]{diaz}Diaz, A.J., Khomenko, E., Collados, M., 2014,
A\&A 564, A97

\bibitem[Fryxwell et al. (1991)]{fryxwell} Fryxwell, B., Arnett, D., \& Mueller, E.,
1991, ApJ, 367, 619

\bibitem[Hester et al. (1996)]{hester} Hester, J.J., Stone, J.M., Scowen, P.A. et al.,
1996, ApJ, 456, 225

\bibitem[Hillier (2017)]{hillier} Hillier, A., 
2017, Rev. Mod. Plasma Phys. 2:1

\bibitem[Innes et al. (2012)]{innes} Innes, D.E., Cameron, R.H., Fletcher, L. et al., 
2012, A\&A, 540, L10

\bibitem[Isobe et al. (2005)]{isobe05} Isobe, H., Miyagoshi, T., Shibata, K., Yokoyama, T., 2005
Nature, 434, 478

\bibitem[Isobe et al. (2006)]{isobe06} Isobe, H., Miyagoshi, T., Shibata, K., Yokoyama, T.,
2006, PASJ, 58, 423

\bibitem[Jones et al. (2005)]{jones} Jones, T.W., \& De Young, D.S.,
2005, ApJ, 624, 586

\bibitem[Jun \& Norman (1996)]{junnorman} Jun, B.I., \& Norman, M.L., 
ApJ, 624, 586

\bibitem[Jun et al. (1995)]{jun} Jun, B.I., Norman, M.L., \& Stone, J.M., 
1995, ApJ, 465, 800

\bibitem[Kruskal \& Schwartzschild (1954)]{kruskalschwartzschild} Kruskal, M., Schwartzschild, M.,
1954, Proc. R. Soc. Lond. Ser. A, 223, 348

\bibitem[Kulkarni et al. (2008)]{kulkarni} Kulkarni, A.K., Romanova, M.M.,
2008, MNRAS 386, 673

\bibitem[Livescu (2004)]{livescu} Livescu, D.,
2004, Phys. Fluids, 16, 118

\bibitem[Maatsumoto \& Masad (2013)]{M&M} Maatsumoto, J. \& Masad, Y.,
2013, ApJ, 772, L1

\bibitem[Mishra et al. (2018)]{mishra18} Mishra, S.K., Singh, T., Kayshap, P., Srivastava, A.K., 2018, ApJ, 874, 57

\bibitem[Mishra \& Srivastava (2019)]{mishra19} Mishra, S.K. \& Srivastava, A.K., 2019, ApJ, 856, 86

\bibitem[Okamoto et al. (2007)]{okamoto} Okamoto, T.J., Tsuneta, S., Berger, T.E. et al. 2007,
Science, 318, 5856

\bibitem[O'Neill et al. (2009)]{oneill} O'Neill, S.M., De Young, D.S., \& Jones, T.W.,
2009, ApJ, 694, 1317

\bibitem[Rayleigh (1900)]{rayleigh} Rayleigh, L., 
190, \textit{Scientific Papers}, vol. II (Cambridge University Press, Cambridge) p.200

\bibitem[Ribeyre et al. (2004)]{ribeyre} Ribeyre, X., Tikhonchuk, V.T., \& Boquet, S.,
2004, Phys. Fluids, 16, 4661

\bibitem[Robinson et al. (2004)]{robinson} Robinson, K., Dursi, L.J., Ricker, P.M., et al.,
2004, ApJ, 601, 621

\bibitem[Ruderman et al. (2014)]{ruderman+t+b} Ruderman, M.S., Terradas, J., \& Ballester, J.L.,
2014, ApJ, 785, 110

\bibitem[Ruderman (2017)]{ruderman17} Ruderman, M.S.,
2017, Sol. Phys., 292, 47

\bibitem[Ruderman et al. (2017)]{ruderman+b+k} Ruderman, M.S., Ballai, I., Khomenko, E., \& Collados, M.,
2017, A\&A 609, A23

\bibitem[Ryutova et al. (2010)]{ryutova} Ryutova, M., Berger, T., Frank, Z., \& Title, A.,
2010, Sol Phys., 267, 75

\bibitem[Shivamoggi (1982)]{shivamoggi} Shivamoggi, B.K.,
1982,  Phys. Fluids, 25, 911

\bibitem[Taylor (1950)]{taylor} Taylor, G.I.,
1950, RSPSA, 201, 192

\bibitem[Terradas et al. (2008)]{terradas08} Terradas, J., Arregui, I., Oliver, R., \& Ballester, J.L.,
ApJ Letters, 678, 2

\bibitem[Terradas et al. (2012)]{terradas} Terradas, J., Oliver, R., \& Ballester, J.L.,
2012, A\&A, 541, A102

\bibitem[Vandervoort(1961)]{vandervoort} Vandervoort, P.O.,
1961, ApJ, 66, 56

\bibitem[Vickers et al. (2018)]{vickers} Vickers, E., Ballai, I., \& Erd\'elyi, R., 
2018, Sol. Phys., 293, 10, 139

\bibitem[Wang et al. (1983)]{wang} Wang, Y.M., \& Nepveu, M., 1983, A\&A, 118, 267 


\end{thebibliography}
\end{document}